\documentclass[12pt]{article}
\usepackage{epsfig}

\def\beq{\begin{equation}}
\def\eeq#1{\label{#1}\end{equation}}
\def\eeqn{\end{equation}}

\def\beqa{\begin{eqnarray}}
\def\eeqa#1{\label{#1}\end{eqnarray}}
\def\eeqan{\end{eqnarray}}

\let\bar=\overbar

\def\Dslash{\not{\hbox{\kern-4pt $D$}}}
\def\dslash{\not{\hbox{\kern-2pt $\del$}}}

\def\msb{{\bar{\ssstyle M \kern -1pt S}}}

\def\Title#1{\begin{center} {\Large {\bf #1} } \end{center}}

\begin{document}

\Title{Transverse sphericity of charged particles in minimum bias p-p collision in ALICE at the LHC}

\bigskip\bigskip


\begin{raggedright}  

{\it Antonio Ortiz Velasquez\index{Reggiano, D.}\\
(for the ALICE Collaboration) \\
Instituto de Ciencias Nucleares, Universidad Nacional Aut\'onoma de M\'exico\\
Circuito Exterior, Ciudad Universitaria, Del. Coyoac\'an, C.P. 04510. M\'exico D.F.}
\bigskip\bigskip
\end{raggedright}


\section{Abstract}
The study of the sphericity of primary charged particles in minimum bias proton-proton collisions at $\sqrt{s}=0.9$, $2.76$ and $7$ TeV with the ALICE detector at the LHC is presented. The mean sphericity as a function of multiplicity is reported for events with different hardness. The ALICE results show a growing transverse sphericity with multiplicity for all collision energies, with a steeper rise at low $N_{ch}$. The tendency of the MC generators is opposite at high multiplicity.

\section{Introduction}

The measurements in minimum bias (MB) proton-proton collisions are relevant to provide information about the non-perturbative aspects of QCD and they are the baseline for the correct interpretation of the heavy ion results.

The recent results from the LHC experiments show that the event generators are unable to explain simultaneously different observables. For instance, at $\sqrt{s}=$0.9 TeV, ALICE has reported that for inelastic processes the multiplicities predicted by the different models in general do not reproduce the data, except for PHOJET \cite{phojet} which exhibits the best agreement, but fails in the description of the $p_{T}$ distribution \cite{alice:2,alice:3}.  The main lesson of the LHC is that an important effort has to be done in order to understand all the observations. Toward this effort, in the present work we investigate the evolution of the transverse sphericity with the multiplicity.

\section{Transverse Sphericity Analysis in ALICE}

At hadron colliders the event shapes are restricted to the transverse plane in order to avoid the bias from the boost along the beam axis \cite{banfi2004}. In this work the transverse sphericity is measured in the acceptance $|\eta|\leq0.8$, for events with more than two tracks ($p_{T}\geq0.5$ GeV/c). The observable is defined as follows:

\begin{equation}
S_{T} \equiv \frac{2\lambda_{2}}{\lambda_{2}+\lambda_{1}}
\label{eqSt}
\end{equation}

where: $\lambda_{1}>\lambda_{2}$ are the eigenvalues of the transverse momentum matrix:

\begin{displaymath}
\mathbf{S_{xy}^{L}} = \frac{1}{\sum_{i} p_{Ti}}\sum_{i}
	\frac{1}{p_{Ti}}\left(\begin{array}{cc}
{p_{xi}}^{2}      &  p_{xi} p_{yi} \\
p_{xi} p_{yi} &  {p_{yi}}^{2} 
	\end{array} \right)        
\end{displaymath}

By construction, an isotropic event has sphericity close to 1, and the ``jetty'' events have $S_{T} \rightarrow 0$.

The mean transverse sphericity ($\langle S_{T} \rangle$) was measured as a function of the multiplicity at three center of mass energies. Each MB sample (``bulk'') was split in ``soft'' ($p_{T}^{\mathtt{max}}<2$ GeV/c) and ``hard'' ($p_{T}^{\mathtt{max}}\geq2$ GeV/c) events.  The method used to correct the measured sphericity {\it vs.} multiplicity for efficiency, acceptance, and other detector effects is similar to one used in the mean $p_{T}$ analysis \cite{alice:3}.

The overall number of analyzed events is $\sim$40 million events for each $\sqrt{s}$ = 7 and 2.76 TeV, and 3.6 millions for  0.9 TeV. The beam-related background rejection is performed using the VZERO detector consisting of two forward scintillator hodoscopes.  The MB trigger required a hit in one of the VZERO counters or in the Silicon Pixel Detector (SPD). In addition a coincidence between the signals from two beam pickup counters, one on each side of the interaction region, indicating the presence of passing bunches was required \cite{alice:2}. For the measurement of the charged $\langle S_{T} \rangle$ the relevant detectors are the Time Projection Chamber (TPC) and the Inner Tracking System (ITS), both located in the central barrel of ALICE \cite{alice-ppr1}.

\section{Results}

\begin{figure*}
\begin{center}
\begin{tabular}{ccc}
\resizebox{0.33\textwidth}{!}{\includegraphics{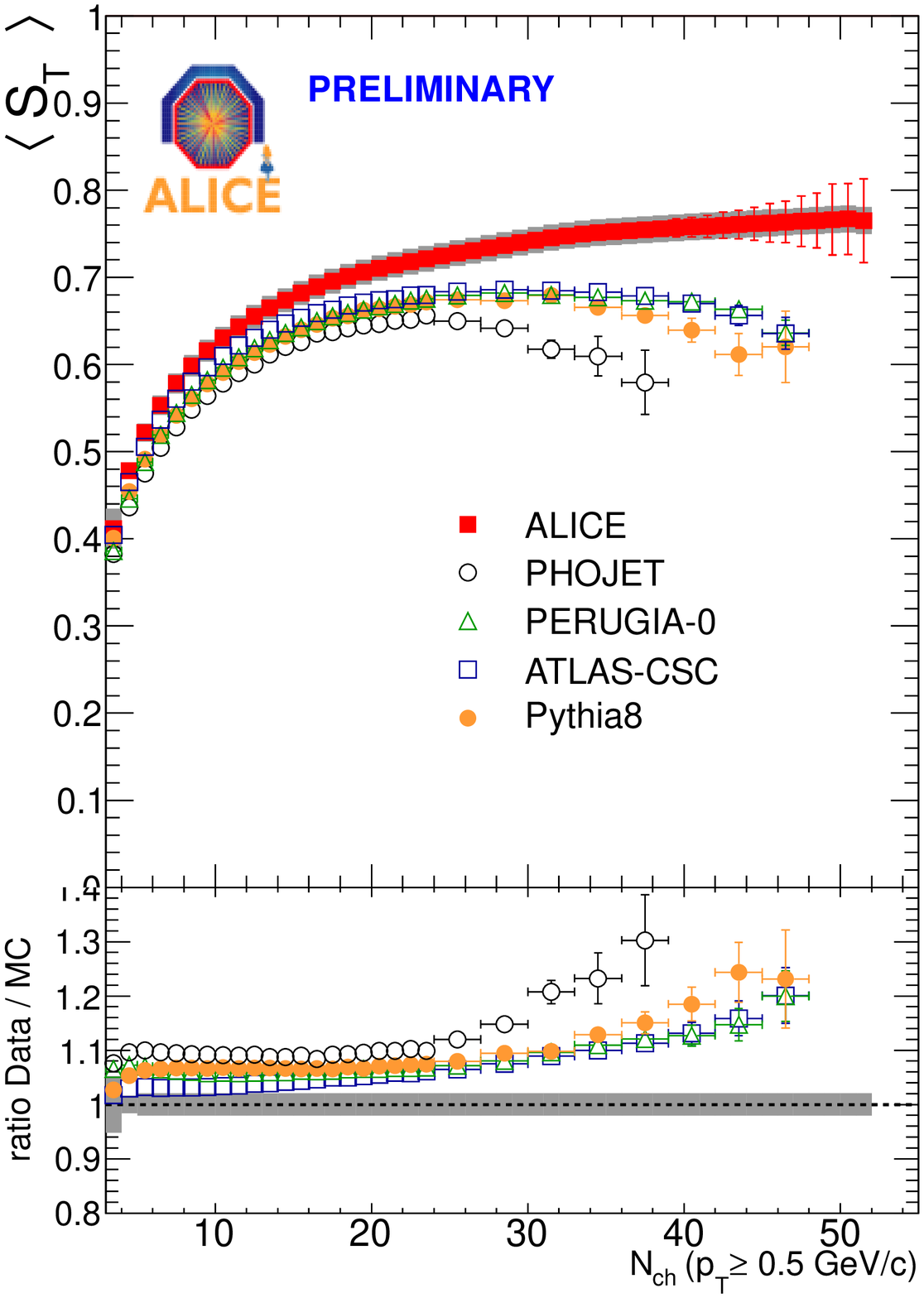} } &
\resizebox{0.33\textwidth}{!}{\includegraphics{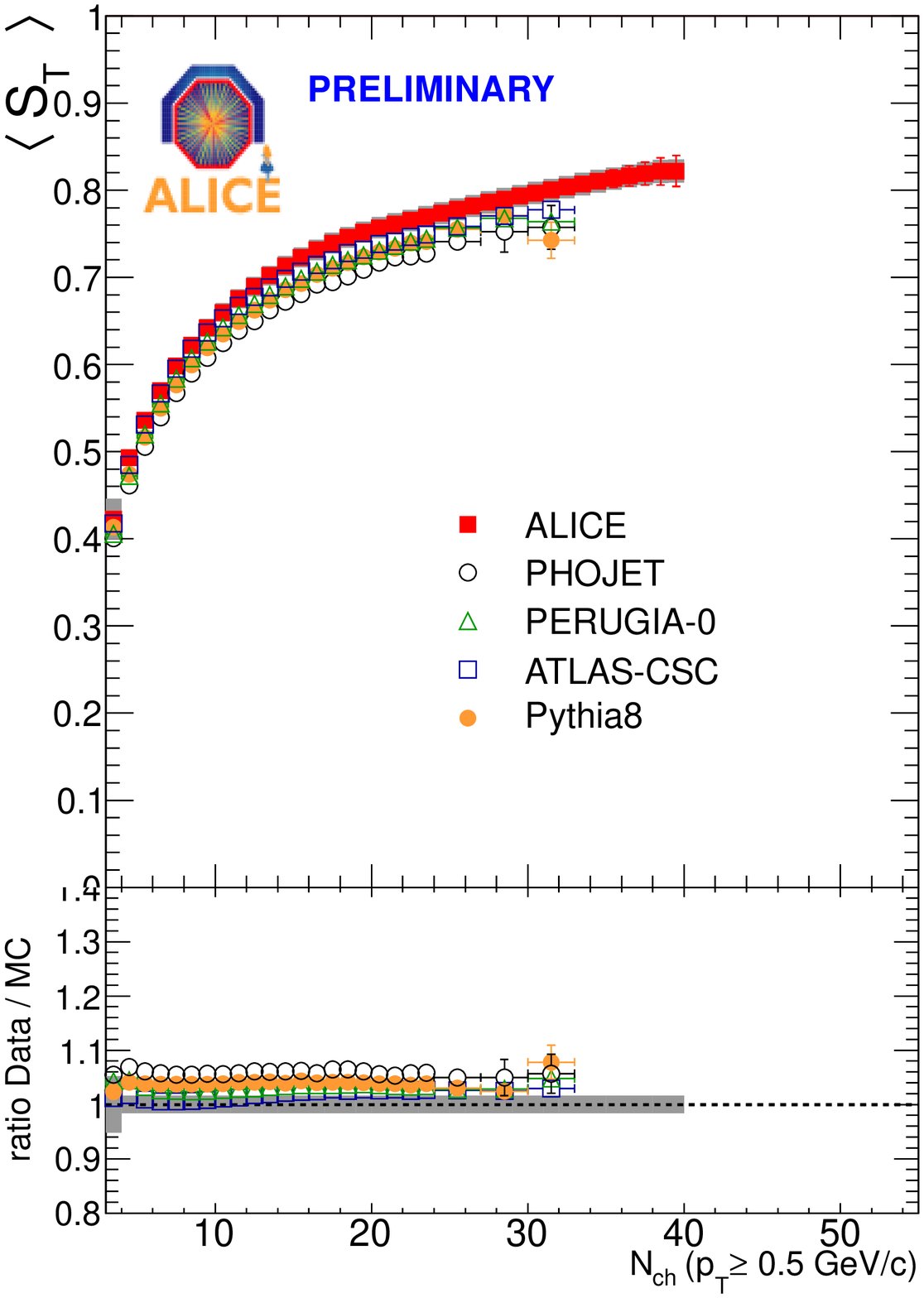} } &
\resizebox{0.33\textwidth}{!}{\includegraphics{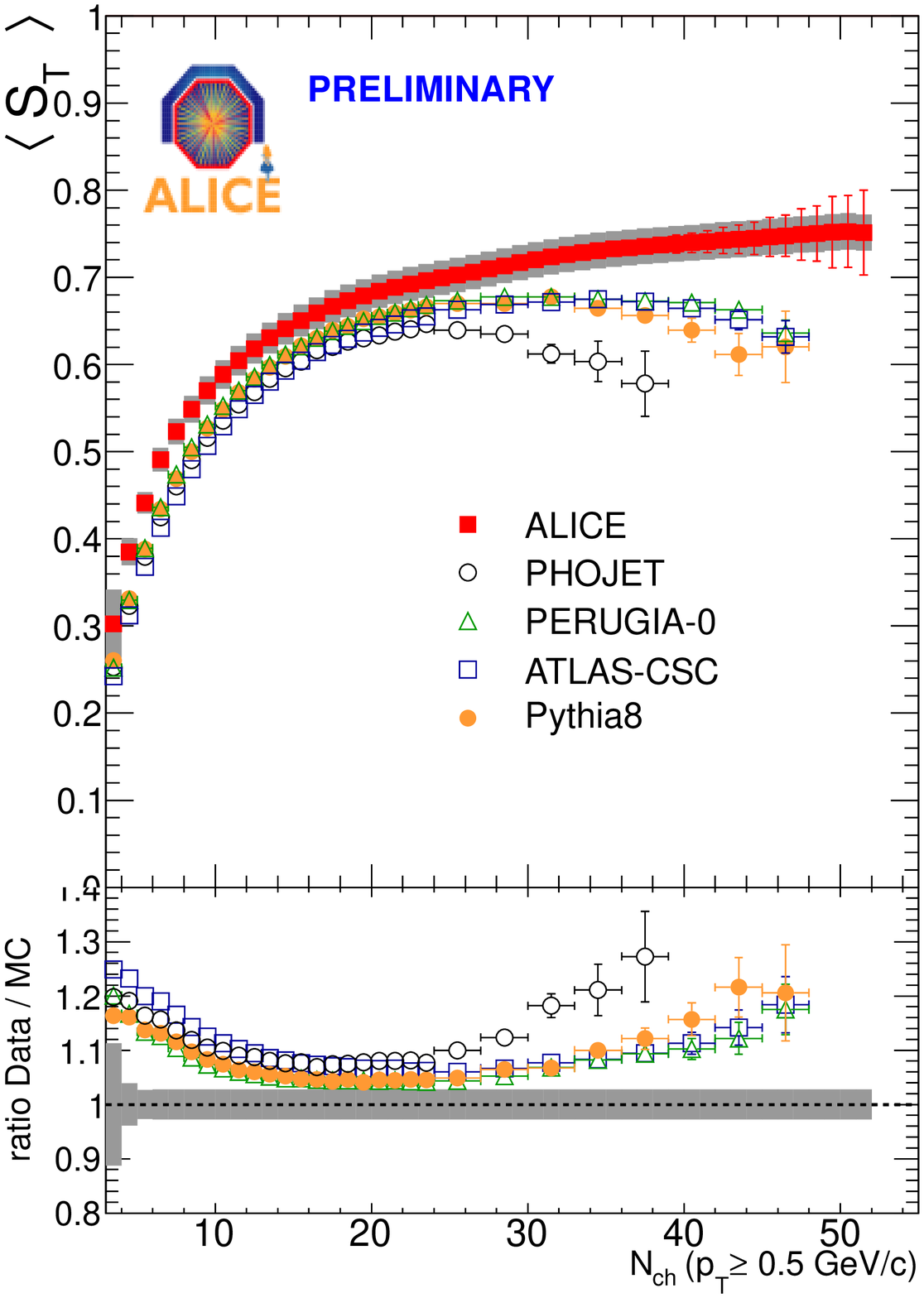} } 

\\
(a) & (b) & (c)
\\
\end{tabular}
\caption{ Mean transverse sphericity as a function of charged particle multiplicity for pp collisions at $\sqrt{s}=7$ TeV. The statistical errors are displayed as error bars and the systematic uncertainties as the shaded area. The results are shown for the different event classes: (a) ``bulk,'' (b) ``soft'' and (c) ``hard.''}
\label{alice7:1}
\end{center}
\end{figure*}

The mean transverse sphericity as a function of multiplicity is shown in Fig. \ref{alice7:1} for MB interactions at $\sqrt{s}=7$ TeV, the results are compared with PYTHIA6 \cite{pythia6} version 6.421 (tunes: ATLAS-CSC \cite{atlascsc} and PERUGIA-0 \cite{perugia}), PHOJET \cite{phojet} version 1.12 and PYTHIA8  (untuned parameters \cite{pythia8}) version 8.145. The ALICE data are plotted with the systematic uncertainties, in particular the contribution due to pile-up is $<0.3\%$. For ``bulk'' events, the sphericity in data is steadily rising with multiplicity suggesting a more isotropic distribution of tracks in azimuth than the models. One sees that the general agreement between models is better for ``soft'' events while for the ``hard'' ones the disagreement is up to $\sim 20 \%$ at low and high multiplicity.  

\begin{figure*}
\begin{center}
\resizebox{0.9\textwidth}{!}{ \includegraphics{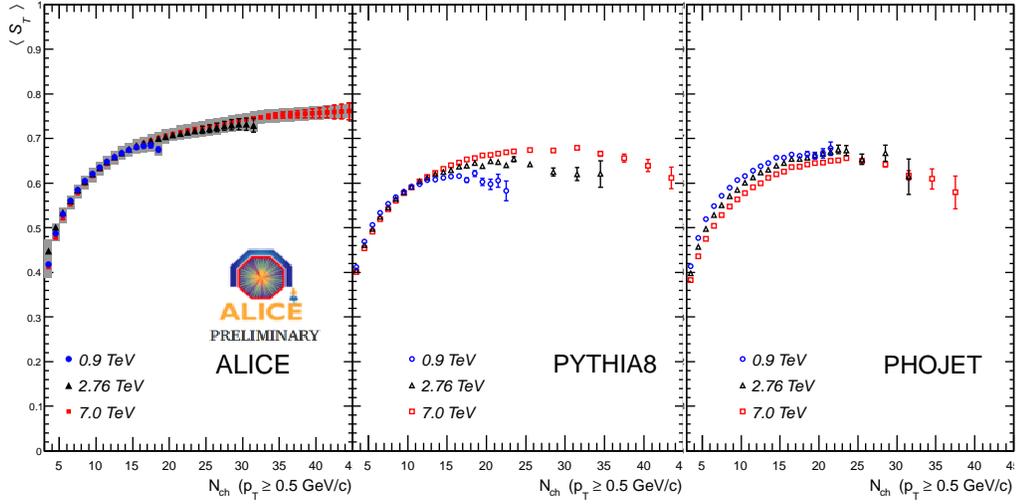} 
  }\vspace{0.2cm}
  \caption{Mean sphericity versus multiplicity for bulk events. Three samples at different energies were analyzed: $\sqrt{s}=$ 0.9, 2.76 and 7 TeV. ALICE data are compared with PYTHIA8 and PHOJET.}
\label{sten1}       
\end{center}
\end{figure*}

On the other hand, Fig. \ref{sten1} shows a comparison of the observable at $\sqrt{s}=0.9$, 2.76 and 7 TeV for ``bulk'' events. The ALICE data are compared with PYTHIA8 and PHOJET. 

\section{Conclusions}

A study of the transverse sphericity of charged particles in MB p-p collisions at $\sqrt{s}=$0.9, 2.76 and 7 TeV was reported. At high multiplicity ($N_{ch}\geq30$), the measured $\langle S_{T} \rangle$ differs by about $20\%$ with respect to different models: ALICE data exhibit a growing trend while the models show a decreasing behavior. The observed effect can not be explained by any systematic uncertainty. On the other hand, the functional form of $\langle S_{T} \rangle$ as a function of $N_{ch}$ is the same at all three energies in the overlapping region. The MC generators show a different behavior. It would be interesting to include the sphericity parameter for the MC tuning.

\section{Acknowledgments}
This work has been done with the close collaboration of Guy Paic. I acknowledge useful discussions and suggestions of Andreas Morsch and Eleazar Cuautle. The work has been supported by DGAPA-UNAM under PAPIIT grants 
IN116008, IN115808 and IN116508 as well as by CONACyT grant 103735, 101597.  Several stays at CERN were supported by the CERN-UNAM mobility program and by the red FAE.

\end{document}